\documentstyle[osa,preprint]{revtex}  % DON'T CHANGE
\begin{document}                % INITIALIZE - DONT CHANGE
\newcommand{\beq}{\begin{equation}}
\newcommand{\enq}{\end{equation}}
\newcommand{\bee}{\begin{eqnarray}}
\newcommand{\ene}{\end{eqnarray}}
\newcommand{\bem}{\begin{mathletters}}
\newcommand{\enm}{\end{mathletters}}
\newcommand{\non}{\nonumber}

\def\btt#1{{\tt$\backslash$#1}}
\def\BibTeX{\rm B{\sc ib}\TeX}
%\begin{document}
%\draft
%\preprint{HEP/123-qed}

\title{Double resonant processes in $\chi^{(2)}$ nonlinear periodic media}

\author{V. V. Konotop$\dag$ and V. Kuzmiak$\ddag$}

\address{ $\dag$
Departamento de F\'{\i}sica and Centro de F\'{\i}sica da Mat\'eria
Condensada,  Universidade de Lisboa, Complexo Interdisciplinar,
Av. Prof. Gama Pinto, 2, Lisbon, P-1649-003 Portugal
\\
$\ddag$ Institute of Radio Engineering and Electronics, Czech
Academy of Sciences, Chaberska 57, 182 51 Prague 8,Czech Republic}
%

%\date{\today}
\maketitle

\begin{abstract}

In a one-dimensional periodic nonlinear $\chi^{(2)}$ medium, by
choosing a proper material and geometrical parameters of the
structure, it is possible to obtain two matching conditions for
simultaneous generation of second and third harmonics. This leads
to new diversity of the processes of the resonant three-wave
interactions, which are discussed within the framework of slowly
varying envelope approach. In particular, we concentrate on the
fractional conversion of the frequency $\omega \rightarrow (2/3)
\omega$. This phenomenon  occurs by means of intermediate energy
transfer to the first harmonic at the frequency $\omega/3$ and can
be controlled by this mode. By analogy the same medium allows
"nondirect" second harmonic generation, controlled by the cubic
harmonic. Propagation of localized pulses in the form two coupled
bright solitons on first and third harmonics and a dark soliton on
the second harmonic is possible.

\end{abstract}

\pacs{PACS numbers: 42.79.Dj}

\narrowtext

\section{Introduction}

In recent years a concept of photonic band gap (PBG) materials
attracted much attention because of their potential applications
in many technical and scientific areas
\cite{review1,joannopoulos1}.  The most of the theoretical studies
of PBG materials focused on the problem of the solution of the
dispersion relation that contains an information on the
propagation of electromagnetic waves through periodic dielectric
structures in the linear regime, when the dielectric constant is
independent of the  field intensity. On the other hand, it is well
known that the introduction of an intensity-dependent refractive
index can significantly change the transmission properties of a
medium.  A considerable activity devoted to the investigation of
the phenomena associated with the propagation of the
electromagnetic waves in nonlinear media provided a bulk of
evidence that PBG structures possessing nonlinearity imply the
existence of many interesting and useful phenomena which can occur
via nonlinear interaction and therefore are believed to be equally
fruitful as PBG systems based on the linear regime
\cite{winful1,eggle,winful2,sterke1}.  For example, it has been
found that the existence of the PBG in which linear
electromagnetic effects are absent does not exclude the
possibility of nonlinear wave propagation \cite{winful2,sterke1}.
It has been demonstrated that for certain values of the input
power the radiation with the frequency in the stop gap can be
transmitted due to the excitation of nonlinear solitary waves
which may provide practical way of coupling large amounts of
optical energy into otherwise unpenetrable photonic bandgap. If
the dielectric has either a positive or negative Kerr coefficient
the transmission properties of a medium are dramatically changed.
For instance, the range of the forbidden wavelengths can be
altered interactively by the nonlinear medium response and this
effect can induce an intensity-dependence which has important
application in optical bistable switching.\cite{scalora1}
Recently, the study of bistability in periodic $\chi^{(2)}$
materials has been reported in Ref.\ \onlinecite{buryak}.

On the other hand, the possibility of multiplication or division
by an integer of a frequency of an electromagnetic wave in a
nonlinear medium, in particular the second and third harmonic
generation is of great practical importance
\cite{yab1,blomb,ZI,mart,fejer,steel,dowling1}. In this paper we
examine the possibility of the fractional transformation of the
frequency which occurs in the photonic crystal represented by 1D
periodic multilayer system possessing $\chi^{(2)}$ nonlinearity.
This phenomenon as well as any scheme of nonlinear conversion must
include phase matching mechanism which is usually achieved by use
of an appropriate birefringent nonlinear crystal or by employing
some kind of the phase matching which corrects the relative phase
between the interacting waves by modulating the sign or magnitude
of the nonlinear coefficient on the scale of the coherence length
while the linear part of the refractive index is continuous. In
contrast, the phase matching in photonic crystals is accomplished
by periodic modulation of the linear refractive
index\cite{blomb,ZI} or by introducing a defect mode \cite{mart}.
The enhanced gain in such structures comes about as a result of
the interplay of the high electromagnetic mode density density in
the vicinity of the band edge due to the modified boundary
conditions\cite{purcell} and strong confinement of both the pump
and higher harmonic signals. Simultaneously, low group velocity
leads to larger interaction of the interacting waves. The process
of the {\em fractional frequency conversion} which we consider in
this paper represents a particular case of three wave interaction
in systems with $\chi^{(2)}$ nonlinearity. Namely, we show that a
simultaneous combination of $\omega =  3\omega - 2\omega$ process
and the second harmonic generation in a 1D periodic nonlinear
medium possessing only $\chi^{(2)}$ nonlinearity results in the
output frequency conversion $\omega \rightarrow (2/3)\omega$.

In the Section II propose 1D periodic PBG structure in which the
resonant conditions for simultaneous SHG and THG and derive a set
of coupled-mode equations which allow to determine the evolution
of the intensities of the interacting modes. In Section III we
analyze stability of the second harmonic of a constant amplitude
in terms of the system of the evolution equations and determine
the conditions which define regions of stability. In Section IV a
stationary process is considered in more detail when the initial
energy is arbitrarily distributed among the modes for
phase-matched waves. In Section V we identify different types of
the evolution of the intensities in terms of the potential
governing newtonian particle and present the results of numerical
simulations which correspond to the different regimes. In Section
VI a particular solitary wave solution is presented in the case
when the coefficients $\gamma_1$ and $\gamma_3$ are of the same
order. In Section VII we summarize the results and discuss both
the material the structural parameters which affect the efficiency
of the processes considered and conditions under which these
phenomena can be described in the frame of the parabolic
approximation.

\section{Statement of the problem}

\subsection{Resonant conditions and phase-matching in 1D Photonic Band
Gap Structure}

Let us consider a TE polarized plane wave incident on a periodic
medium with the dielectric perimittivity
$\hat{\epsilon}(x,\omega)$, $\hat{\epsilon}(x,\omega) =
\hat{\epsilon}(x+L,\omega)$, where $L$ is a period of the
structure.  We assume that the medium possesses a $\chi^{(2)}$
nonlinearity which is also periodic with the period $L$:
$\chi^{(2)}(x) =  \chi^{(2)}(x+L)$. The inclusion of the isotropic
second order nonlinearity leads to the nonlinear polarization of
the form
\begin{equation}
\label{pnl} P_{NL}(x,t) =   \chi^{(2)}(x) E^2(x,t).
\end{equation}

The Maxwell equations for the TE-polarized wave ${\bf E} =
(0,E,0)$ propagating along the $x$-axis, i.e. $E\equiv E(x,t)$,
lead to the wave equation
\begin{eqnarray}
\label{princ} -c^2\frac{\partial^2 E(x,t)}{\partial
x^2}+\frac{\partial^2}{\partial
t^2}\int_{-\infty}^{\infty}\epsilon(x,t-t')E(x,t')\,dt'
=
-4\pi\frac{\partial^2}{\partial t^2} P_{NL}(x,t)
\end{eqnarray}
where
\begin{equation}
\epsilon(x,t-t') =  \delta(t-t') + 4\pi \chi(x,t-t')
\end{equation}
and $\chi(x,t)$ is the linear permittivity [while
$\hat{\epsilon}(x,\omega)$ is the Fourier transform of
$\epsilon(x,t)$].

Specifically, we will consider a periodic multilayered structure
consisting of two different components denoted as $a$ and $b$ and
fabricated from the materials possessing $\chi^{(2)}$ nonlinearity
(as the  matter of fact it is enough that only one of the slabs is
nonlinear). Then we define the filling fraction $f =  a/(a+b)$ as
an important {\em geometrical} parameter which characterizes the
stack.

Let us assume that the filling fraction $f$ and the dielectric
permitivities of the two types of layers $\epsilon_{a}$,
$\epsilon_{b}$ are chosen in such a way that there exist three
frequencies $\omega_j\equiv\omega(q_j)$  $(j = 1,2,3)$ such that
\begin{equation}
\label{res1} \omega_3 =  3\omega_1+\Delta\omega_3,\,\,\,\,\,\,
\omega_2 =  2\omega_1+\Delta\omega_2
\end{equation}
where $|\Delta\omega_j|\ll\omega_1$ and
\begin{equation}
\label{res2} q_3 =   3q_1+Q_1,\,\,\,\,\,\,\,q_2 =    2q_1+Q_2,
\end{equation}
where  $Q_j$ ($j = 1,2)$ are vectors of the reciprocal lattice,
are satisfied. In such a system one has two resonant conditions
for the SHG and third harmonic generation (THG) fulfilled
simultaneously (if $\Delta\omega_2 =  \Delta\omega_3 = 0$ one has
the exact phase matching). We notice that one of the possibilities
to satisfy the mentioned resonant conditions which we call the
double resonance hereafter, is to choose all $\omega_j$ bordering
stop gaps such that (i) all $\omega_j$ belong to the center of the
Brillouin zone (BZ) or (ii) $\omega_1$ and $\omega_3$ are at the
boundary and $\omega_2$ is at the center of the BZ \cite{KK}.
However, both possibilities are not very interesting from the
point of view of practical applications, since it is essential to
have energy transfer among modes in space, while the group
velocities of modes bordering a stop gap are zero. Therefore we
choose the frequencies  $\omega_j$ to be shifted towards allowed
bands and therefore possessing non-zero group velocities $v_j$.

We have shown that such resonant conditions can be achieved in the
1D periodic structure consisting of the alternating slabs of
Al$_{0.1}$Ga$_{0.9}$As and InSb characterized by the dielectric
constant $\epsilon_{GaAs} =  10.97$ and $\epsilon_{InSb} = 16.4$
at $\lambda =  2\mu m$. Both Al$_{0.1}$Ga$_{0.9}$As and InSb
possess the second order nonlinearity with $\chi^{(2)} =
1.68\cdot 10^{-10}$ m/V  and $\chi^{(2)} =    1.84\cdot 10^{-10}$
m/V, respectively. In Fig.~\ref{f1}(a) we present a photonic band
structure for 1D periodic system consisting of the alternating
layers of Al$_{0.1}$Ga$_{0.9}$As and InSb when the filling
fraction $f =    0.1$. It is demonstrated that in such a system
the resonant conditions given by Eqs. (\ref{res1}) and
(\ref{res2}) can be satisfied for both wave vectors and
frequencies of the fundamental and the final states of the second
and third harmonics. The band structure shown in Fig. 1(a) is
determined by using standard plane wave method \cite{plihal} for
an 1D infinite system and the frequencies are expressed in
normalized units. In order to correlate our simulations with the
experimentally accessible systems we choose the lattice constant
$a =  0.7334 \mu m$. The frequencies are taken such that the
fundamental signal represented by a laser source operating at $2
\mu m$  corresponds to the frequency close to the band gap edge of
the third lowest band and the second and third harmonic signals
with the wavelengths $1 \mu m$ and $0.667 \mu m$, respectively,
correspond to the frequencies from the vicinity of the 7. and 10.
lowest gap, respectively. In Fig. 1(a) we indicate the dispersion
curve at $\lambda =   2 \mu m$ which corresponds to the
fundamental signal by a solid line for six lowest bands (as it
would be without material dispersion). The broken lines refer to
the 7. lowest band at $\lambda =  1 \mu m$, which corresponds to
the second harmonic and to the 10. lowest band at $\lambda =
0.667 \mu m$ which belongs to the third harmonic signal. In
evaluating the photonic band structure for the fundamental and the
final higher harmonic states we include the material dispersion in
the range of the frequencies considered by substituting the
tabulated values of the dielectric constants \cite{palik} at
$\lambda =   2 \mu m$: $\epsilon_{AlGaAs} =  10.97$,
$\epsilon_{InSb} = 16.4$; at $\lambda = 1 \mu m$:
$\epsilon_{AlGaAs} = 11.86$, $\epsilon_{InSb} =  18.23$ and at
$\lambda =  0.5 \mu m$: $\epsilon_{AlGaAs} =  17.77$,
$\epsilon_{InSb} = 13.32$. The phase matching between the
interacting waves was accomplished by varying of the filling
fraction of the slabs and choosing the components with suitable
material parameters that constitute a geometrical configuration in
which the resonant conditions for both nonlinear processes are
satisfied. By taking into account the material dispersion of
AlGaAs and InSb associated with the wavelengths corresponding to
the fundamental signal, the second and third harmonic we have
found that the phase matching between the fundamental and the
second harmonic and between the fundamental and the third harmonic
occurs for the opposite propagating waves. In what follows without
restriction of generality the geometry is taken such that the
fundamental (i.e. having the lowest frequency wave) is always
forward-propagating. In Fig.~\ref{f1}(b) we show in detail the
dispersion curves in the extended scheme in the range of the wave
vectors and the frequencies in the vicinity of the first lowest
band where the regions of the curves having positive and negative
derivatives correspond to forward- and backward-traveling waves,
respectively. The phase matching between the forward-traveling
fundamental and the second harmonic waves is determined by the
intersection of the two curves which represent the fundamental
signal indicated by the solid line and a dashed curve which
represents the second harmonic divided by a factor 2. We show that
the exact phase matching between the fundamental wave and the
second harmonic is possible when the wave vector $q =  q_{SHG} =
0.468$ and the frequency $\omega a/ 2 \pi c =  0.3667$, while the
exact phase matching between the fundamental wave and the third
harmonic is possible when the wave vector $q = q_{THG} = 0.471$
and the frequency $\omega a/ 2 \pi c =  0.3680$. The phase
matching between the forward-traveling fundamental and third
harmonic wave is determined by the intersection of the two curves
which represent the fundamental signal indicated by the solid line
and a dash-dotted curve which represents the third harmonic
divided by a factor 3. Since in the system considered the {\it
exact phase matching} between the fundamental and second harmonic
and the fundamental and third harmonic occur for the wave vectors
$q_{SHG} =  0.468$ and $q_{THG} =  0.471$ that are not identical,
we introduce parameters $\Delta q_{SHG}$, $\Delta q_{THG}$,
$\Delta \omega_{SHG}$, and  $\Delta \omega_{THG}$, that
characterize the conditions under which {\it approximate phase
matching} occurs. Namely we choose the wave vector $q =  0.47$ at
which the frequency of the fundamental signal is $\omega a/ 2 \pi
c = 0.3677$ and corresponding frequency of the second harmonic
divided by factor 2 yields $\omega a/ 2 \pi c =   0.3667$ while
the frequency of the third harmonic divided by the factor 3 yields
$\omega a/ 2 \pi c =  0.3680$. Then the mismatch in frequency
between the fundamental and the second the harmonic wave at $q =
0.47$ is $\Delta\omega_2 =  \Delta \omega_{SHG} =  -0.001$ in
respect to the wave vector $q =   0.468$ at which an exact phase
matching between the fundamental wave and the second harmonic
occurs. The mismatch in frequency between the fundamental and
third harmonic wave at $q =  0.47$ is $\Delta\omega_3 =  \Delta
\omega_{THG} = 0.0003$ in respect to the wave vector $q =  0.471$
at which an exact phase matching between the fundamental wave and
the third harmonic takes place.

\subsection{Evolution equations}

The wave evolution in the processes of higher harmonic generation
is well described in terms of the so-called  envelope function
approach\cite{sterke1}. In the system described above one has to
take into account three resonant waves, i.e. to express the
electric field in the form
\begin{equation}
E = \sum_{j=1}^{3}A_j\phi_j(x)e^{i\omega_j t} + c.c.
\end{equation}
where $\phi_j(x)$ ($j=1,2,3$) are orthogonal and normalized
eigenfunction of the eigenvalue problem
\begin{equation}
\label{eigen} [c^2(d^2/dx^2) +
\hat{\epsilon}_0(x;\omega_j)\omega_j^2]\phi_j(x) =  0
\end{equation}
$A_j(x)$ ($j=1,2,3$) is  a slowly varying, compared with
$\phi_j(x)e^{i\omega_j t}$, amplitude of $j$-th mode, and $c.c.$
stands for  the complex conjugate. Using the conventional and well
elaborated procedure  (see e.g.\cite{sterke1}) one arrives at the
system of equations
\begin{eqnarray}
\label{e1a} \frac{i}{\omega_1}\left(\frac{\partial A_1}{\partial
t} + v_1\frac{\partial A_1}{\partial x}\right) +
\gamma_3\bar{A}_2A_3e^{i(\Delta\omega_3-\Delta\omega_2)t} +
2\gamma_1\bar{A}_1A_2e^{i\Delta\omega_2t}=  0
\end{eqnarray}
\begin{eqnarray}
\label{e1b} \frac{i}{\omega_2}\left(\frac{\partial A_2}{\partial
t}+ v_2\frac{\partial A_2}{\partial x}\right) +
\gamma_3\bar{A}_1A_3e^{i(\Delta\omega_3-\Delta\omega_2)t} +
\bar{\gamma}_1A_1^2e^{-i\Delta\omega_2 t}=  0
\end{eqnarray}
\begin{eqnarray}
\label{e1c} \frac{i}{\omega_3}\left(\frac{\partial A_3}{\partial
t}+ v_3\frac{\partial A_3}{\partial x}\right) +
\bar{\gamma}_3A_1A_2e^{-i(\Delta\omega_3-\Delta\omega_2)t} = 0
\end{eqnarray}
where the coefficients are given by
\begin{equation}
\label{gamma1} \gamma_1 =
2\pi\int_{0}^{L}\chi^{(2)}(x;-\omega,2\omega)\bar{\phi}_1(x)
\bar{\phi}_1(x)\phi_2(x)\,dx
\end{equation}
\begin{equation}
\label{gamma3} \gamma_3 =
2\pi\int_{0}^{L}\chi^{(2)}(x;-\omega,3\omega)\bar{\phi}_1(x)
\bar{\phi}_2(x)\phi_3(x)\,dx
\end{equation}
$L$ is a length of the stack, and we assume the following symmetry
relations \cite{LL}:
\[\chi^{(2)}(x;\omega_1,\omega_2)=
\chi^{(2)}(x;-\omega_3,\omega_2)=
\chi^{(2)}(x;\omega_1,-\omega_3)\]
are taken into account and bar denotes the complex conjugation.

We note that the system of Eqs. (\ref{e1a})-(\ref{e1c}) is derived
for a finite structure subject to periodic boundary conditions
\cite{sterke1}. The outcomes, however, can be extended to an
infinite structure by computing the limit $L\to \infty$. The
difference between the results obtained for finite and infinite
structures  is estimated to be of $O(L^{-1})$. Bearing this in
mind in what follows analytical results are derived for an
infinite (or semi-infinite) structure and for numerical estimates
of the coefficients $\gamma_j$ a finite structure are used.
Another important comment to be made is that we consider
(\ref{e1a})-(\ref{e1c}) also on a semi-line $x\geq 0$. Formally to
do this one should use proper eigenfunctions $\phi_j(x)$ of the
boundary value problem (rather than conventional Bloch functions).
For special types of solutions, however, (say possessing a
symmetry with respect to origin) this leads to a negligible error
which very roughly can be estimated to be of order of $\lambda/l$,
where $\lambda$ is a length of the carrier wave and $l$ is a
characteristic scale of spatial variations of $A_j$ (in other
words to be of order of the small parameter of the structure).

It is worth to point out that in the case of pulse propagation (or
more precisely when $\lim_{t\to\pm\infty}A_j(x,t) = 0$) one of the
integrals of the evolution system (\ref{e1a})-(\ref{e1c}) is given
by
\beq \label{MR} v_1W_1(x)+v_2W_2(x)+v_3W_3(x) =  0 \enq
where
\beq \label{energy} W_j(x) =
\int_{-\infty}^{\infty}|A_j(x,t)|^2\,dt. \enq

\section{Stability of the second harmonic}

One of the solutions of the Eqs. (\ref{e1a})-(\ref{e1c}) which is
of a special importance for the next consideration is that of
possessing the second harmonic having a constant amplitude the and
first and third harmonics having zero amplitude. We refer such a
solution as {\it background}, hereafter. Then, considering an
infinite system, the question about stability of such a state
appears. We consider this problem  within the framework of  system
of the equations (\ref{e1a})-(\ref{e1c}) for slowly varying
amplitudes which are rescaled as follows
\beq \label{amplit} a_1 =
\sqrt{\frac{2}{w_2}}A_1\exp(-i\delta\omega_1t),\qquad a_2 =
A_2\exp(i\varphi_1),\qquad a_3 =
\frac{\Gamma_3}{\Gamma_1}\sqrt{\frac{2}{w_2}}A_3
\exp[i(-\delta\omega_3t+\varphi_3 + \varphi_1)]\,. \enq
In the case of perfect phase matching ($\delta\omega_2 =
\delta\omega_3 = 0$), we have
\beq \label{s1a} i\left( \frac{\partial a_1}{\partial
\tau}+\frac{\partial a_1}{\partial
\xi}\right)+\bar{a}_2a_3+2\bar{a}_1a_2 =   0\,, \enq
\beq \label{s1b} i\left( \frac{1}{w_2}\frac{\partial a_2}{\partial
\tau}+\sigma_2 \frac{\partial a_2}{\partial
\xi}\right)+\bar{a}_1a_3+a_1^2 =  0\,, \enq
\beq \label{s1c} i\left( \frac{1}{w_3}\frac{\partial a_3}{\partial
\tau}+\sigma_3 \frac{\partial a_3}{\partial \xi}\right)+\mu a_1a_2
=  0\,. \enq
Here $\tau =  \omega_1\Gamma_1 t$,  $\xi =
(\omega_1\Gamma_1/v_1)x$, $\sigma_j =   $sign$(v_j)$ $w_j =
|v_j|/v_1$, $\Gamma_j =  |\gamma_j|$, $ \mu =
\frac{3v_1\Gamma_3^2}{|v_3|\Gamma_1^2}$, and $\phi_j =
\arg(\gamma_j)$. (Recall that in  accordance with our convention
$v_1>0$.)

We are interested in the stability properties of a solution of the
Eqs. (\ref{s1a})-(\ref{s1c}) as follows: $a_{1,3} = 0$ and $a_2 =
\rho\exp[i(\omega \tau-k\xi)]$, where $\omega = \sigma_2 w_2 k$
and $\rho$ is a positive constant playing the role of the
amplitude of the background. Following the conventional procedure
we consider perturbations $\alpha_j$ of the mentioned solution,
i.e. we substitute $a_n = \alpha_n\exp[ni(\omega\tau-k\xi+\phi)]$
for ($n = 1,3$) and $a_2 = (\rho +
\alpha_2)\exp[2i(\omega\tau-k\xi+\phi)]$, where
$|\alpha_j|\ll\rho$ into the Eqs. (\ref{s1a})-({\ref{s1c}). Next
we linearize the system of equations with respect to the
$\alpha_j$ and represent $\alpha_j =
\alpha_{0j}\exp[i(\Omega\tau-K\xi)]$ where $\alpha_{0j}$ are
constants. Then equations for $\alpha_1$ and $\alpha_3$ are
singled out and we obtain the dispersion relation $f(\Omega,K) =
0$ for small excitations against the background. The complex roots
of this equation indicate instability. After straightforward
analysis, one obtains the conditions for $f(\Omega,K) = 0$ having
no complex roots. They read
\beq \label{stab1} k^2\left(1-\frac{v_2}{v_1}\right)^2>4 \enq
\beq \label{stab2} 3\left(\Bigg|k\left(1-\frac{v_2}{v_1}\right)
\Bigg|+2\right) \Bigg|k  \left(1-\frac{v_2}{v_3}\right) \Bigg|<
\mu \enq
\beq \label{stab3} (v_1-v_2)(v_2-v_3)>0 \enq

The first consequence comes out from (\ref{stab1}): only a
background in a form of a plane wave, i.e. with $k\neq 0$, is
stable. The respective boundary conditions to
(\ref{s1a})-(\ref{s1c}) read
\beq \label{bound} \lim_{\xi\to\infty}a_{1,3} =
0,\,\,\,\,\,\,\lim_{\xi\to\infty}a_2= \rho e^{i(k\xi-w_2\tau)}
\enq
An important consequence of (\ref{stab3}) is that the stable
second harmonic can be achieved in a structure where phase
matching takes place for waves with "ordered" group velocities:
either $v_1>v_2>v_3$ or  $v_1<v_2<v_3$. In particular, an unstable
case is when the first and third harmonics are forward
propagating, and the second harmonic propagates in the opposite
direction.

As it has been mentioned above in a real configuration it is
almost impossible to avoid phase mismatches $\Delta\omega_j$.
They, however, result in the same terms of the evolution equations
as the "rotation" of the background. Hence phase mismatch can
either enhance or suppress the instability of a static background.
Below it will be convenient to consider $\omega_2$ as a reference
frequency, and define $\delta\omega_1 =  -\Delta\omega_2/2$ and
$\delta\omega_3 = \Delta\omega_3-\frac 32 \Delta\omega_2$ as a
mismatch of the frequencies of the first and third harmonic,
respectively. Then, considering the Eqs. (\ref{s1a})-(\ref{s1c})
subject to the boundary conditions (\ref{bound}), taking into
account the phase mismatch, and introducing new dependent
variables $\tilde{a}_n =  a_n\exp[i\frac{n}{2}(k\xi-w_1\tau)]$,
one arrives at the system of equations
\beq \label{s2a} i\left( \frac{\partial \tilde{a}_1}{\partial
\tau}+\frac{\partial \tilde{a}_1}{\partial
\xi}\right)+\bar{\tilde{a}}_2\tilde{a}_3
+2\bar{\tilde{a}}_1a_2+\tilde{\nu_1}\tilde{a}_1 = 0\,, \enq
\beq \label{s2b} i\left(\frac{1}{w_2}\frac{\partial
\tilde{a}_2}{\partial \tau} +\sigma_2\frac{\partial
\tilde{a}_2}{\partial \xi}\right)+
\bar{\tilde{a}}_1\tilde{a}_3+\tilde{a}_1^2 = 0\,, \enq
\beq \label{s2c} i\left(\frac{1}{w_3}\frac{\partial
\tilde{a}_3}{\partial \tau} +\sigma_3\frac{\partial
\tilde{a}_3}{\partial \xi}\right)+ \mu
\tilde{a}_1\tilde{a}_2+\tilde{\nu}_3\tilde{a}_3 = 0 \enq
where $\nu_j = \frac{\Delta\omega_j}{\omega_1}\frac{\sigma_j}
{\Gamma_1 w_j}$, $\tilde{\nu}_j = \delta_j-\nu_j$, $\delta_1 =
\frac 12 k(1-w_2)$, and $\delta_3 = \frac{3k(w_3-w_2)}{2w_3}$ .

It follows from the above discussion, that in the particular case
when it is possible simultaneously satisfy the two conditions
$\delta_j = \nu_j$ the case of quasi-phase matching is
mathematically equivalent to the case of exact phase matching with
the background in a form of a plane wave. Thus, one can associate
an effective wave vector of the background with a phase mismatch
and if this wave vector belongs to the stability region then one
can speak about stabilization of the background by the phase
mismatch.

\section{Stationary process}

Let us consider in more detail a stationary  process when $A_2$
does not depend on slow time while the temporal dependence of
$A_j$ ($j = 1,3$) is harmonic $\sim e^{-i\delta\omega_j t}$. Then
the system of the evolution equations (\ref{e1a})-(\ref{e1c}) can
be rewritten in the form
\begin{eqnarray}
\label{a1a} i\frac{d a_1}{d\xi} +
\bar{a}_2a_3+2\bar{a}_1a_2-\nu_1a_1 =    0
\end{eqnarray}
\begin{eqnarray}
\label{a1b} i\sigma_2\frac{da_2}{d\xi} +\bar{a}_1a_3+a_1^2 =    0
\end{eqnarray}
\begin{eqnarray}
\label{a1c} i\sigma_3\frac{da_3}{d\xi}+  \mu a_1a_2-\nu_3 a_3 =
0
\end{eqnarray}

First we concentrate on the case of exact phase matching:
$\Delta\omega_2 = \Delta\omega_3 = 0$ ($\nu_1 =  \nu_3 =  0$).
Then system of the Eqs. (\ref{a1a})-(\ref{a1c}) allows a
particular exact solution describing the energy transfer from the
first and third modes to the second one\cite{KK}. That takes
place, however, only for a specific relation between the
amplitudes of the first and  third modes. In the present paper we
focus  on a more generic situation when the input energy is
distributed among modes arbitrarily. In particular, we are
interested in the case when initially the second mode is not
excited and in the effect of each of modes on the energy transfer
to the second harmonic.

To this end we note that the system of the Eqs.
(\ref{e1a})-(\ref{e1c}) possesses two integrals:
\begin{eqnarray}
\label{e2} N =  |a_1|^2 + 2\sigma_2 |a_2|^2 + \frac{3\sigma_3}{
\mu}|a_3|^2
\end{eqnarray}
[$dN/d\xi = 0$,  it substitutes the relation (\ref{MR}) in the
case of monochromatic waves]  and
\begin{eqnarray}
\label{e3} H =  a_1^2\bar{a}_2 + \bar{a}_1^2a_2+ a_1a_2\bar{a}_3 +
\bar{a}_1\bar{a}_2a_3
\end{eqnarray}
($dH/d\xi = 0$).

For the next consideration it is convenient to separate real and
imaginary parts of the amplitudes, $a_1 = q_1 + ip_1$, $a_2 = q_2
+ i\sigma_2p_2$, $a_3 = q_3 + i\sigma_3 \mu p_3$. Then one can
consider the system of the Eqs. (\ref{a1a})-(\ref{a1c}) as a
dynamical one generated by the Hamiltonian
\begin{eqnarray}
\label{hamilt} H =    \sigma_2\sigma_3 \mu
q_1p_2p_3+2\sigma_2q_1p_1p_2 - \sigma_{2}p_1p_2q_3
+q_1q_2q_3+\sigma_3 \mu p_1q_2p_3 +q_1^2q_2-p_1^2q_2
\end{eqnarray}
with the canonical Poisson brackets and conjugated variables $q_j$
and $p_j$. A remarkable property of this system is that the
Hamiltonian admits another representation:
\begin{equation}
\label{hamilt1} H =    q_2\frac{d p_2}{d\xi} - p_2\frac{d
q_2}{d\xi}
\end{equation}
which gives the relation between the amplitude and the phase of
the second harmonic.

In a generic case, when $H\neq 0$ the dynamics is characterized by
permanent energy exchange among modes.  In what follows, however,
we will be especially interested in situations when all energy is
concentrated in one or two of higher harmonics. This  can happen
only if  $H =  0$: indeed, if the field of the first or the second
harmonic is zero, then $H$ is zero  as well.

The peculiarity of the frequency conversion among modes in
periodic media is that the modes involved in the process can be
either co-propagating or counter-propagating. Naturally, the
behavior of the system strongly depends on the case under
consideration, as we already have seen on example of the
stability. We can distinguish four different cases: (i) forward
propagating waves ($\sigma_2=\sigma_3=1$), (ii) backward
propagating second harmonic ($\sigma_2=-1$ and $\sigma_3=1$),
(iii) backward propagating third harmonic ($\sigma_2=1$ and
$\sigma_3=-1$), and (iv) backward propagating second and third
harmonics ($\sigma_2=\sigma_3=-1$). Notice that the last case
occurs in the structure depicted in Fig.~\ref{f1}.

In order to study the dynamical system for $q_j$ and $p_j$ we
observe that it follows from (\ref{hamilt1}) that when $H = 0$ one
obtains $q_2/p_2 =  \cot\theta_2 $ where $\theta_2$ is the phase
of the second harmonic (hereafter we speak about the phases of
$a_j$) which does not depend on the $\xi$ and thus is an integral
of motion. On the other hand, by computing $p_2$ from
(\ref{hamilt}) we obtain
\beq \label{intC} \cot\theta_2 =  \frac{p_1q_3- \mu
q_1p_3-2q_1p_1}{q_1q_3+ \mu p_1p_3+q_1^2-p_1^2} \enq
The last formula allows one to express the functions  $q_1$,
$p_1$, $q_3$, and $p_3$ through the other ones.

The next consideration is devoted to the case when the second
harmonic has phase shift $\pi/2$ with respect to the first and
third ones: \beq \label{theta} \theta_2 = \pm \pi/2. \enq This
means that $q_2\equiv 0$ and $p_2\neq 0$ at any point in space.
Then by introducing a new variable \beq \label{taup} X_p =
\sigma_2\int_0^{\xi} p_2(\xi')\,d\xi ' \enq
one can linearize the dynamical system  which now takes the form
\beq \label{sys01} \frac{dq_1}{dX_p}-q_3+2q_1 = 0 \enq
\beq \label{sys02} \frac{dq_3}{dX_p}+\sigma_3 \mu q_1 = 0 \enq
\beq \label{sys03} \frac{dp_1}{dX_p}-\frac{q_3p_1}{q_1} = 0 \enq

The first two equations are linear and thus are trivially resolved
with respect to $q_1$ and $q_3$:
\beq \label{solut1a} q_1 = C_1e^{\omega_+X_p} + C_2e^{\omega_-X_p}
\enq
\beq \label{solut1b} q_3 =  -C_1\omega_-
e^{\omega_+X_p}-C_2\omega_+ e^{\omega_-X_p} \enq
where
\bee \label{omega} \omega_{\pm} =  \left\{\begin{array}{ll} -1\pm
\sqrt{1-\sigma_3 \mu}&\,\,\,\,\, \mbox{if $\sigma_3 \mu<1$}\\
-1\pm i \sqrt{\sigma_3 \mu-1}&\,\,\,\,\, \mbox{if $\sigma_3
\mu>1$}
\end{array}\right.
\ene

Then for $p_1$ we obtain
\beq \label{solut1c} p_1 =
p_{01}e^{2X_p}\left(C_1e^{\omega_+X_p}+ C_2e^{\omega_-X_p}\right)
\enq
The constants $C_1$, $C_2$, and $p_{01}$ are determined by the
amplitude of the field at the input of the structure.

\section{Fractional frequency conversion}

Let us look for stationary points of the Eqs.
(\ref{sys01})-(\ref{sys03}), i.e. for solutions tending to
constant at $\xi\to\infty$. In the case when they exist,
stationary points correspond to situation where the energy of the
electromagnetic field is either distributed among modes or
concentrated in one of them and energy exchange among modes does
not occurs any more. Then one can distinguish two cases: (i)
$X_p\to\pm \infty$ what happens when $\pm \sigma_2 p_2^{(st)}>0$,
and (ii)  $X_p\to X^{(st)} = const.$ what happens when $p_2^{(st)}
=  0$, where $p_2^{(st)} =  \lim_{\xi\to \infty}p_2(\xi)$. The
first situation corresponds to the case when the energy is
converted to the second harmonic and the second one corresponds to
the case when the energy output of the second harmonic is zero. In
this section we concentrate on the former case.

It follows from  the Eqs. (\ref{solut1a}), (\ref{solut1b}), and
(\ref{solut1c}) that in the case of forward propagating third
harmonic ($\sigma_3 = 1$) a bounded nontrivial solution can exists
only if $p_{01} = 0$. Moreover, a convergent solution of the Eqs.
(\ref{sys01}), (\ref{sys02}) can exist only when
$\sigma_2p_2^{(st)} > 0$ what corresponds to the $q_1^{(st)} =
q_3^{(st)} =  0$. In physical terms this conclusion means that the
energy initially distributed among three modes is totally
transferred during the process into the second harmonic. The
necessary condition for such process requires the first, $a_1$,
and third $a_3$, harmonics to be forward propagating waves and
have phase difference either $0$ or $\pi$ while the second one,
$a_2$ has a relative phase shift $\pm\pi/2$ [the actual phases of
the modes are defined by the last conditions and the
relations(\ref{amplit})].

In order to understand which parameters of the input signal result
in stationary points we observe that $X_p$ as a function of $\xi$
can be treated as a coordinate of an effective newtonian particle
governed by the equation (the case $\sigma_3 = -1$ is included in
the next consideration, as well)
\beq \label{eqx1} \frac{d^2X_p}{d\xi^2} = -\frac{\partial
U(X_p)}{\partial X_p} \enq
The potential $U(X_p)$ is easily computed from the equation for
$q_2$ and (\ref{solut1a}), (\ref{solut1b}) in terms of elementary
functions for arbitrary value of  $ \mu$. It has, however, a
rather cumbersome form to be represented here. Instead, we mention
that the dynamics of the effective particle essentially depends on
whether (a) $0 < \sigma_3 \mu < 1$, (b) $\sigma_3 \mu > 1$ or (c)
$\sigma_3 \mu < 0$ and $ \mu > 1$. In order to simplify the
discussion of these cases  we assume that there is no input second
harmonic. This means that initially (i.e. at $\xi=   0$) $X_p =
0$, $dX_p/d\xi = 0$ i.e. the energy of the effective particle is
determined by the point of intersection of the energy curve with
$U$-axis,  and that the type of the motion depends on the value of
the potential energy at $X_p = 0$. In Fig.~\ref{f2} we display
three typical configurations of $U(X_p)$ which correspond to the
above cases. They represent different kinds of motion:
Fig.~\ref{f2}(a) corresponds to motion of the particle to the
right which ends up in the steady state motion. This type of
motion means that $X_p\to = \sigma_2p_2^{(st)}\xi$ and thus
corresponds to the total energy transfer from the first and third
harmonics to the second one. In Fig.~\ref{f2}(b) the particle is
initially in a potential well and thus undergoes oscillations
around the local minimum of the effective potential. This type of
motion, corresponds to periodic energy exchange among modes.
Finally, Fig.~\ref{f2}(c) corresponds to exponentially growing
solution which indicates the instability within the framework of
the parabolic approximation.

The first two types of the behavior described above are
represented in Figs.~\ref{f3}(a) and \ref{f3}(b), where we show
the results of the numerical simulations obtained by solving the
dynamical equations (\ref{a1a})-(\ref{a1c}) for slowly varying
amplitudes $A_j$ ($j = 1,2,3$). We display the evolution of the
intensities $|A_j|^2$ ($j =  1,2,3$) as a function on $\xi$. In
Figs.~\ref{f3}(a), (b) we have chosen the same initial values of
the amplitudes of the input signal and the same structure
parameters as those used in Figs.~\ref{f2}(a) and \ref{f2}(b). In
Fig.~\ref{f3}(a) energy transfer from the first and the third
harmonic to the second one is shown. The time of the energy
transfer is inversely proportional to the square root of the field
intensity which is a direct consequence of the evolution
equations. Taking into account that the amplitude of the first
harmonic is relatively small, this situation can be viewed as
frequency down-conversion -- $\omega_3 \rightarrow (2/3)\omega_3$
-- in the case when the input energy is dominated by the pump
signal with the frequency $\omega ( = \omega_3)$. Then after the
transition time the energy is transferred into the mode with the
fractional frequency $\frac 23\omega$ $( =   \omega_2)$. The
fractional down-conversion is observed even when the first
harmonic is negligibly small (but nonzero) at the input. The
existence of the first harmonic in this process is fundamental.
Indeed, $A_3\equiv$ const., $A_1\equiv 0$, and $A_2\equiv 0$ is a
solution of the dynamical system (\ref{e1a})-(\ref{e1c}) and thus
the first harmonic is necessary in order to initiate the process
of the down conversion. Moreover, at small enough intensities of
the first harmonic we have observed local amplification of the
first harmonic.

The case which corresponds to the periodic exchange of the energy
among modes is shown in Fig.~\ref{f3}(b). Here we are close to the
minimum of the potential energy and the interaction among modes is
rather weak (the second harmonic is not zero but has a very small
amplitude). Almost the whole energy of the electric field is
concentrated in the third harmonic. However it is possible to find
parameters at which interactions among modes are strong and even
locally the whole energy is concentrated in one or two modes.

As it is evident, by a proper choice of initial parameters in the
case of a "potential " depicted in Fig.~\ref{f2}(b) an effective
particle can be placed outside the potential well. Then the
fractional frequency conversion occurs. This situation is given in
Fig.~\ref{f3}(c). Finally, in Fig.~\ref{f3}(d) it is shown the
exchange of the energy among modes in the presence of mismatch.
Although the second harmonic is still generated (i.e. fractional
down conversion occurs) at specific distances which are placed
almost periodically one observes enhancement of the first
harmonic.

Returning to the example of a real structure described in Section
2 one cannot provide frequency conversion to one of the
frequencies because of strong instability (this instability
however is an artifact of the parabolic approximation: direct
numerical simulations of this case will be reported elsewhere).
Meantime, when the third harmonic has na amplitude much higher
than the first and second harmonic, one can achieve a stable
propagation of the modes, which, however,  manifest very weak
interaction. This situation is shown in Fig.~\ref{f4}. The
stability of the picture represented is very sensitive to the
relative phases and the amplitudes of the waves.

\section{One solitary wave solution}

Let us now return to evolution equations (\ref{s2a})-(\ref{s2c}).
As it has been mentioned, the phenomenon described above is
essentially based on the geometry of the structure. As a
consequence the effective nonlinear coefficients $\gamma_1$, and
$\gamma_3$ can be of different orders. In particular, one can
achieve the relation $|\gamma_1|\gg|\gamma_3|$. Then the equations
(\ref{s2a})-(\ref{s2c}) in the absence of phase mismatch are
reduced to the conventional system for the SHG which is integrable
by means of the inverse scattering technique \cite{solit}. Another
integrable case occurs when $|\gamma_3|\gg|\gamma_1|$. Then the
Eqs. (\ref{s2a})-(\ref{s2c}) are reduced to the system describing
decay of the pump wave, which is the third harmonic, and the back
process: $A_3\leftrightarrow A_1 + A_2$. A characteristic feature
of the last process is that the frequency of the three waves are
related by (\ref{res1}).

In the case when $\gamma_1$ and $\gamma_3$ are of the same order
it is still possible to find a particular solitary wave solution
having the form of the coupled bright solitons of the first and
third harmonics and a dark soliton associated with the second
harmonic. It reads
\beq \label{solit1} \tilde{a}_1 = \frac{\alpha_{\pm}\beta
u_1}{\cosh(\beta\zeta)} \enq
\begin{equation}
\label{solit2} \tilde{a}_2 = i\frac{\alpha_{\pm}^2\beta u_1^2}{1-
\mu\alpha_{\pm}^2u_1} \left(i\frac{\tilde{\nu}_3}{\beta u_1} +
\tanh (\beta\zeta)\right)
\end{equation}
\begin{equation}
\label{solit3} \tilde{a}_3 = \frac{ \mu \alpha_{\pm}^3\beta u_1^2}
{1- \mu\alpha_{\pm}^2u_1} \frac{1}{\cosh (\beta\zeta)}
\end{equation}
Here
\[\zeta = \frac{(\xi-v\tau)v_2}{v_2-vv_1},
\]
$v$ is a velocity of the pulse,
\[u = \frac{v_3\sigma_3-vv_1}{(1-v)v_3},\qquad
u_1 = \sigma_3-\frac{vv_1}{v_3}\]
\[
\alpha_{\pm}^2 = \frac{\mu (\mu+u)\pm \sqrt{u(u-\mu)}}{\mu
u_1(\mu+3u)}
\]
and the phases of the frequencies of the modes must be chosen to
satisfy the relation
\beq \label{delta} \frac{\tilde{\nu}_3}{\tilde{\nu}_1} =
\frac{(1-\mu \alpha_{\pm}^2 u_1)^2}{\alpha_{\pm}^2u_1
(\mu\alpha_{\pm}^2u_1-2)}. \enq

\section{Discussion and Conclusion}

Within the slowly varying amplitude approximation we derived a set
of coupled mode equations which describe the evolution of the
intensities of the resonant waves. We used these equations to
study double resonant processes in $\chi^{(2)}$ nonlinear periodic
media in which two matching conditions for the generation of the
second and third harmonic are simultaneously satisfied. We show
that such conditions can be achieved in one-dimensional photonic
band gap structure by choosing a proper combination of geometrical
and material parameters.

Speaking about possibilities of experimental fabrication of
structures with double phase matching we notice that it is more
difficult to match parameters, than say for the second harmonic
generation only. In particular it is intriguing when actual
experimental values of the refractive index are taken into
account. Meantime it seems that there are no restriction for
possibility to find a proper geometry almost for any $\chi^{(2)}$
material. Another possible technological difficulty could be a
lattice mismatch of the layers. In this context it is worth noting
that a periodic structure is rather flexible in the sense that it
allows inclusion of new components which being much thinner than
the main nonlinear slabs do not affect the phenomenon of frequency
conversion. Namely in the case when chosen layers possess large
lattice mismatch, slabs of the third kind having width much
smaller than $a$ or $b$ could be included between nonlinear
layers.

It has been shown that the use of double resonances allows one to
obtain difference frequency generation. A particular example is
$\omega\to(\frac 23)\omega$. This however is not direct conversion
but it occurs with participation of the mode with the frequency
$\omega/3$.

Parabolic approximation which for a given scaling results in  a
dynamical system is generated by Hamiltonian $H$. The non-zero
Hamiltonian corresponds to the permanent energy exchange among the
modes, while $H = 0$ corresponds to the case when the energy can
be (subject to definite conditions) concentrated in one or two
higher harmonics (that is the case of fractional conversion). In
processes of the frequency conversion in periodic media the modes
can be either co-propagating or counter-propagating and
consequently the behavior observed is very different. In
particular, one can distinguish the following types of the
dynamics: (a) the total energy is transferred from the first and
third harmonic into the second one, (b) a periodic exchange
between the modes occurs and (c) exponentially growing solution,
which represents instability within the parabolic approximation.
The qualitative picture based on the analysis of the motion of the
effective particle in the potential which indicates different
types regimes is very well reflected by the results obtained from
the numerical simulation carried out to solve dynamical equations.

The evolution equations governing three-wave processes in the
presence of double resonances allow a solitary wave solution in a
form of coupled two bright solitons (they correspond to the first
and third harmonics) and a dark soliton on the second harmonic.

Finally, we also notice that periodic structures, involving
geometrical factors rather than only material properties can
provide us with a large diversity of matching conditions. In
particular one can obtain double resonances as follows [c.f. the
Eqs. (\ref{res1}) and (\ref{res2})]
\begin{equation}
\label{res10} \omega_3 = 2\omega_2+\Delta\omega_3,\,\,\,\,\,\,
\omega_2 = 2\omega_1+\Delta\omega_2\,,
\end{equation}
and
\begin{equation}
\label{res20} q_3 =  2q_2+Q_1,\,\,\,\,\,\,\,q_2 =    2q_1+Q_2\,,
\end{equation}
which result is a different kind of three wave interactions.
Solitonic solutions generated by the respective interactions in
the presence of dispersion have been recently considered in
\cite{malomed}. Naturally, the dispersion can lead also to
solitons in the case of double resonances defined in the Eqs.
(\ref{res1}), (\ref{res2}).

\acknowledgments

VVK is greatful to Boris Malomed for pointing out the reference
\cite{malomed}. VVK acknowledges  support from FEDER and Program
PRAXIS XXI, grant N$^0$ PRAXIS/2/2.1/FIS/176/94. The work has been
partially supported by the bilateral agreement ICCTI - Czech
Academy of Sciences.

\begin{figure} \caption{(a) Photonic band structure of 1D periodic
structure consisting of alternating slabs of
Al$_{0.1}$Ga$_{0.9}$As with $\epsilon_a =  10.97$ and InSb with
$\epsilon_b =  16.4$ at $\lambda =  2 \mu m$.(solid curve) in
which fractional frequency conversion takes place. The broken
curves refer to the 7. lowest band at $\lambda =  1 \mu m$ which
corresponds to the second harmonic and the 10. lowest band at
$\lambda =  0.667 \mu m$ which corresponds to the third harmonic
signal. $q$ is measured in the units $2 \pi/(a+b))$.
(b) The detailed picture of the region of the photonic band
structure shown in Fig. 1a in extended zone scheme in which both
frequencies and the wave vectors that satisfy the resonant
conditions for simultaneous SHG and THG. The solid curve indicates
the calculated dispersion curve near $2 \mu m$, the dashed line
refers to the region near $1 \mu m$ with values divided by a
factor 2 and the dash-dotted line refers to the region near $0.667
\mu m$ with values divided by a factor 3. The exact phase matching
between the forward-travelling fundamental and oppositely
travelling second harmonic occurs when $q = q_{SHG} = 0.468$,
while the exact phase matching between the fundamental wave and
third harmonic is possible for oppositely propagating waves when
$q = q_{THG} = 0.471$.  $q$ is measured in the units $2
\pi/(a+b))$.} \label{f1}
\end{figure}

\begin{figure}
\caption{Examples of the effective potential $U(X_p)$ for the
structure parameters and initial amplitudes of waves as follows:
$\gamma_1 =  82.14 + 78.48i$, $\gamma_3 =  -14.57 - 5.07i$, (a)
$v_1 =  0.5382$, $v_2 =  0.2808$, $v_3 =  0.2433$, $A_1 =  0.13
\cdot 10^9$ V/m, $A_3 =  0.31 \cdot 10^9$ V/m, $A_2 =  0$, $
\mu\approx 0.1225$; (b) $v_1 =  2.382$, $v_2 =  0.2808$, $v_3 =
0.06433$, $A_1=  0.1 \cdot 10^9$ V/m, $A_3 =  -0.9 \cdot 10^9$
V/m, $A_2 =  0$, $ \mu\approx 2.051$; (c) $v_1 =  0.5382$, $v_2 =
0.2808$, $v_3 = -0.2433$, $A_1 = 0.13 \cdot 10^9$ V/m, $A_3 = 0.31
\cdot 10^9$ V/m, $A_2 = 0$, $ \mu\approx 0.1225$.} \label{f2}
\end{figure}

\begin{figure} \caption{Evolution of intensities of the first
(solid line),the second- (broken line) and third-harmonic (dashed
line) signals. (a) the same structure parameters and input
amplitudes as in Figs. 2a; (b) the same structure parameters and
input amplitudes as in Figs. 2b; (c) the same structure parameters
as in Fig. 2b and the input amplitudes $A_1 = 0.1 \cdot 10^9$ V/m,
$A_3 = 0.4 \cdot 10^9$ V/m, $A_2 =  0$; (d)  the same structure
parameters and input amplitudes as in Fig. 2a except the phase
mismatch $\Delta\omega_2\approx-0.002$ and $\Delta\omega_3\approx
0.0009$ was included. The intensities represented are normalized
to $[\chi^{(2)}]^{-2}$ and time is measured in $\omega_1^{-1}$
units.} \label{f3}
\end{figure}

\begin{figure} \caption{Evolution of intensities of the first (solid line),
the second- (broken line) and third-harmonic (dashed line) signals
in a structure depicted in Fig.~\ref{f1}, where the input
amplitudes $A_1 = 0.255\cdot 10^9$ V/m, $A_3 = -1.996\cdot 10^9$
V/m, $A_2 = 0$.} \label{f4}
\end{figure}


\begin{references}

\bibitem{review1} For review see {\it Photonic
Band Gaps and Localization} , edited by C.M Soukoulis (Plenum, New
York, 1993);  the special issue of J. Opt. Soc. Am. B {\bf 10},
1993; the special issue of J. Mod. Opt. {\bf 41} (1994);  {\it
Confined Electrons and Photons},\ NATO ASI Series B Vol. 340,
edited by E. Burstein and C. Weisbuch (Plenum, New York, 1995);
{\it Photonic Band Gap Materials},\ NATO ASI Series E Vol. 315\
edited by C. M Soukoulis (Kluwer, Dordrecht,1996); and in {\it
Microcavities and Photonic Bandgaps: Physics and Applications},\
NATO ASI Series E Vol. 324, edited by J. Rarity and C. Weisbuch
(Kluwer, Dordrecht,1996).

\bibitem{joannopoulos1} J. D. Joannopoulos, R. D. Meade and J. N. Winn,\
{\it Photonic Crystals, Molding the Flow of Light} (Princeton
University Press, Princeton, NJ, 1995).

\bibitem{winful1} H. G. Winful, J. H. Marburger, and E. Garmire,
``Theory of bistability in nonlinear distributed feedback
structures, `` Appl. Phys. Lett. {\bf 35}, 379-382 (1979); L.
Kahn, N. S. Almeida, and D. L. Mills, ''Nonlinear optical response
of superlattices. Multistability and soliton trains,'' Phys.  Rev.
B {\bf 37}, 8072-8081 (1988); V. M. Agranovich, S. A. Kiselev, and
D. L. Mills, ''Optical multistability in nonlinear superlattices
with very thin layers,'' Phys. Rev. B {\bf 44} 10917-10920 (1991).

\bibitem{eggle} B. J. Eggleton, R. E. Slusher, C. M. de Sterke, P.
A. Krug, and J. E. Sipe, ''Bragg grating solitons,'' Phys. Rev.
Lett. {\bf 76}, 1627-1630 (1996).

\bibitem{winful2} H. G. Winful, ``Pulse compression in optical fiber
filters,'' Appl. Phys. Lett.{\bf 46}, 527-529 (1985); W. Chen and
D. L. Mills, ''Gap solitons in nonlinear periodic structures,''
Phys. Rev. Lett. {\bf 58}, 160-163 (1987); D. L. Mills and S. E.
Trullinger, ''Gap solitons in nonlinear periodic structures,''
Phys. Rev. B {\bf 36}, 947-952 (1987).


\bibitem{sterke1} C. M. de Sterke and J. E. Sipe, ''Envelope-function
approach for the electrodynamics of nonlinear periodic
structures,'' Phys. Rev. A {\bf 38}, 5149-5165 (1988).

\bibitem{scalora1} M. Scalora, J. P. Dowling, C. M. Bowden, and
M. J. Bloemer, ''Optical limiting and switching of ultrafast
pulses in nonlinear photonic band gap materials,'' Phys. Rev.
Lett.  {\bf 73}, 1368-1371 (1994); A. Kozhekin and G. Kurizki,
''Self-induced transparency in Bragg reflectors,'' Phys. Rev.
Lett. {\bf 74}, 5020-5023 (1995); M. Scalora, J. P. Dowling, M. J.
Bloemer, and C. M. Bowden, ''The photonic band edge optical
diode'' J. Appl. Phys. {\bf 76}, 2023-2026 (1994); M. Scalora, R.
L. Flynn, S. B. Reinhardt, R. L. Fork, M. J. Bloemer, M. D. Tocci,
J. Bendikson, H. Ledbetter, C. M. Bowden, J. P. Dowling, and R. P.
Leavitt, ``Ultrashort pulse propagation at the photonic band edge:
large tunable group delay with minimal distortion and loss,'',
Phys. Rev. E {\bf 76}, R1078-R1081 (1996).


\bibitem{buryak} A. V. Buryak, I. Towers, and S. Trillo
``Multistability, homoclinic clamping, and chaos in nonlinear
quadratic distributed feedback systems,'' Phys. Lett. A {\bf 267},
319-325 (2000).


\bibitem{yab1} E. Yablonovitch, C. Flytzanis, and N. Bloembergen,
``Anisotropic interference of three-wave and double two-wave
frequency mixing in GaAs,'' Phys. Rev. Lett. {\bf 29}, 865-868
(1972); C. Flytzanis, and N. Bloembergen, ``Infrared dispersion of
third-order susceptibilities in dielectrics: retardation
effects,'' Prog. Quantum Electron. {\bf 7}, 271-300 (1974).

\bibitem{blomb} N. Bloembergen and A. J. Sievers, ''Nonlinear optical
properties of periodic laminar structures,'' Appl. Phys. Lett.
{\bf 17}, 483-485 (1970).

\bibitem{ZI} J. P. van der Ziel and M. Ilegems, ''Optical second
harmonic generation in periodic multilayer
$GaAs-Al_{0.3}Ga_{0.7}As$ structures,'' Appl. Phys.Lett. {\bf 28}
437-439 (1976).

\bibitem{mart} J. Martorell and R. Corbalan, ``Enhancement of second
harmonic generation in a periodic structure with a defect,''
Optics Comm. {\bf 108}, 319-323 (1994);  J. Trull, R. Vilaseca, J.
Martorell, and R. Corbalan, ``Second harmonic generation in local
modes of a truncated periodic structure,''Opt. Lett. {\bf 20},
1746-1748 (1995).

\bibitem{fejer} M. M. Fejer, G. A. Magel, D. H. Jundt, and R. L. Byer,
``Quazi-phase-matched second harmonic generation: tuning and
tolerances,'' IEEE J. Quantum Electron. {\bf 28}, 2631-2654
(1992).

\bibitem{steel} M. J. Steel and C. M. de Sterke, ''Second-harmonic
generation in second-harmonic fiber Bragg gratings,'' Appl. Opt.
{\bf 35}, 3211-3222 (1996); ''Bragg-assisted parametric
amplification of short optical pulses,'' Opt. Lett. {\bf 21}
420-422 (1996); M. Scalora, M. J. Bloemer, A. S. Manka, J. P.
Dowling, C. M. Bowden, R. Viswanathan, and J. W. Haus, ''Pulse
second-harmonic generation in nonlinear one-dimensional, periodic
structures,'', Phys. Rev. A {\bf 56} 3166-3174 (1997).

\bibitem{dowling1} J. P. Dowling, M. Scalora, M. J. Bloemer, and
C. M. Bowden, ``The photonic band edge laser: a new approach to
gain enhancement,'' J. Appl. Phys. {\bf 75} 1896-1899 (1994); M.
Tocci, M. J. Bloemer, M. Scalora, J. P. Dowling, and C. M. Bowden,
``Measurement of spontaneous-emission enhancement near the
one-dimensional photonic band edge of semiconductor
heterostructures,'' Phys. Rev. A {\bf 53} 2799-1783 (1996).

\bibitem{purcell}  E. M. Purcell, ``Spontaneous emission probabilities
at radio frequencies,'' Phys. Rev. {\bf 69}, 681-686 (1946).


\bibitem{KK} V. V. Konotop and V. Kuzmiak, ``Simultaneous second- and
third-harmonic generation in one-dimensional photonic crystals,''
J. Opt. Soc. Am. B {\bf 16}, 1370-1376 (1999).

\bibitem{plihal} M. Plihal and A. A. Maradudin, ``Photonic band structure
of two-dimensional sytems: The triangular lattice,'' Phys. Rev. B
{\bf 44} 8565-8571 (1991).

\bibitem{palik} {\it Handbook of Optical Constants}, edited by
E. D. Palick \ (Academic, New York, 1985).

\bibitem{LL} L. D. Landau and E. M. Lifshitz {\em Electrodynamics of
Continuous Media} (Pergamon Press, 1984).

\bibitem{solit} S. P. Novikov, S. V. Manakov, L. P. Pitaevsky, and
V. E.Zakharov, {\em Theory of Solitons: Inverse Scattering method}
(Consultants Bureau, New York, 1980).

\bibitem{malomed} I. Towers, R. A Sammut, A. V. Buryak, and B. A. Malomed,
``Soliton multistability as a result of double-resonance wave
mixing in $\chi^(2)$ media,'' Opt. Lett., {\bf 24}, 1738-1740
(1999); A. V. Buryak, I. Towers, R. A. Sammut, and B. Malomed in
{\em Nonlinear Guided Waves and Their Applications} OSA Techical
Digest (Optical Society of America, Washington DC, 1998), pp.
64-66.


\end{references}
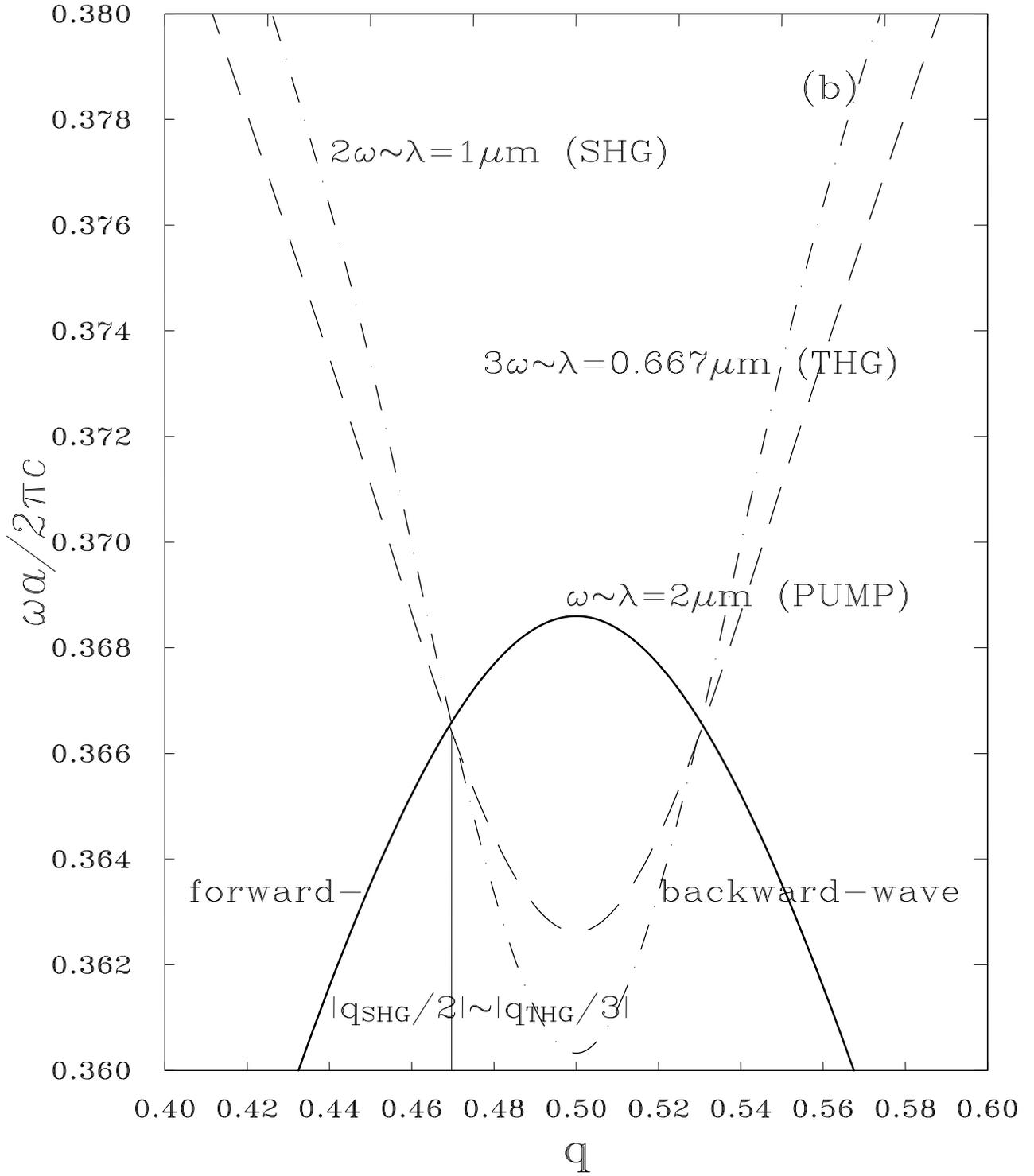
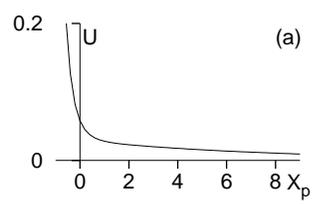
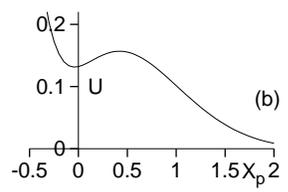
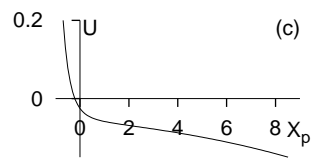
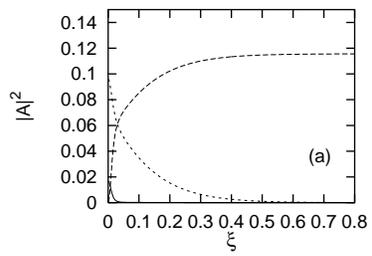
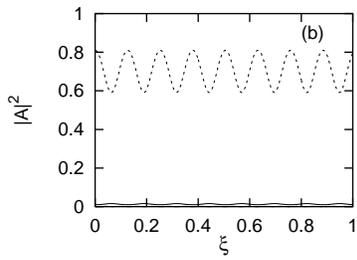
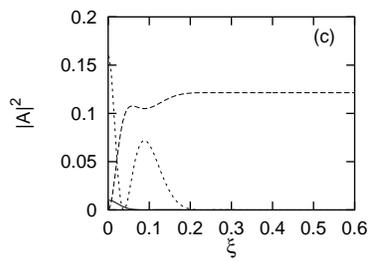
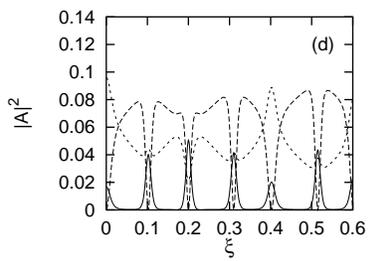
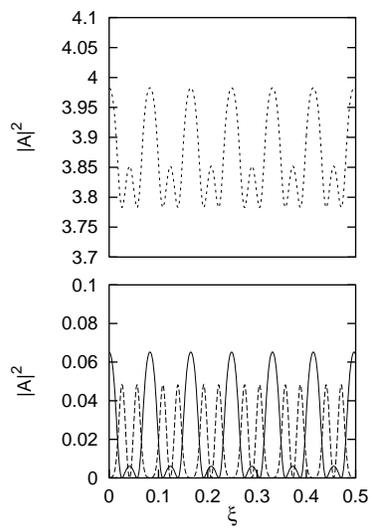
\end{document}